
\input phyzzx

\voffset=24pt
\def\ptslash{\not{\hbox{\kern-2.0pt $p_T$}}}
\def\lslash{\not{\hbox{\kern-0.5pt $l$}}}
\def\rpslash{\not{\hbox{\kern-3.0pt $R_P$}}}
\def\etslash{\not{\hbox{\kern-3.1pt $E_T$}}}
\vfuzz5pt\hfuzz5pt
{\singlespace
\hfill\vbox{\hbox{TIFR/TH/92-29}
            \hbox{UH-511-751-92}
            \hbox{CERN-TH.6613/92}

}
\date{}
\titlepage
\title{\bf Tau Signals of R-parity breaking at LEP 200}
\author{\rm Rohini M. Godbole$^1$, Probir Roy$^2$ and Xerxes Tata$^3$}
\bigskip
\bigskip
{\singlespace
\item{1.} Department of Physics, University of Bombay, Vidyanagari, Bombay
400 098, India and Theory Division, CERN, CH-1211 Geneva 23, Switzerland;
gyan@tifrvax.bitnet

\item{2.} Tata Institute Of Fundamental Research, Homi Bhabha Road, Bombay
400 005, India; probir@tifrvax.bitnet

\item{3.} Department of Physics and Astronomy, University of Hawaii at
Manoa, Honolulu, HI 96822, U.S.A.; tata@uhhepg.bitnet

}
\bigskip

\centerline{\bf Abstract}

\nobreak
The detectability at LEP 200 of explicit $R$-parity breaking by
tau-number $(L_\tau)$ violating operators is considered.  The
assumption of $L_\tau$-violation is motivated by the relative lack of
constraints on such couplings but similar considerations apply to explicit
$L_e$- or $L_\mu$-violation.  The $LSP$, now unstable, and not necessarily
neutral, decays via $L_\tau$-violating modes.  Only signals from the
production and decays of $LSP$ pairs are considered, thereby avoiding any
dependence on the sparticle mass spectrum.  Rather spectacular signals are
predicted: spherical events with $m$ leptons (usually containing at least
one $\tau$) and $n$ jets $(m,n \leq 4)$, the most characteristic of which
are like-sign $\tau\tau$ events.  These signals are enumerated for each
$LSP$ candidate and quantitative estimates are provided for the favoured
case when the $LSP$ is a neutralino.  Other new physics signals, which can
mimic these signatures, are also briefly discussed.

\endpage

\centerline{\fourteenbf 1. Introduction}
\nobreak
Supersymmetry can stabilize the weak scale in the Standard Model (SM) in
an elegant way, provided the superparticles have masses $\lsim 0(1)$ TeV.
The accessibility of this mass range to forthcoming accelerators has made the

phenomenological pursuit of supersymmetry an exciting venture.  Much
effort [1] has, in turn, been expended in that direction over the past few
years.  There exist many model calculations suggesting that, if all
superparticle masses are smaller than about 1 TeV, the lighter charginos
and neutralinos may well have masses below 100 GeV. This means that they
can be pair-produced at LEP 200.  We adopt such an attitude in this paper
and propose some distinctive signals (mostly involving one or more
$\tau$'s along with other charged leptons, jets, $~\etslash $ etc.) to be
looked for at LEP 200 as signatures of a class of $\tau$-number
violating supersymmetric models.  What is special about $\tau$-number --
as will be elaborated below -- is that, among all the conservation laws of
the SM, $\tau$-conservation is the least well-verified [2]. Also, LEP is
ideally suited to search for $\tau-$number violation on account of the
superior $\tau-$detection efficiency offered by its cleaner environment as
compared with a hadron collider.

The main thrust of the effort mentioned above has been within the aegis of
the Minimal Super-Symmetric Model (MSSM) [1].  The MSSM has the particle
content of the SM (but with two Higgs doublets) simply extended by global
$N=1$ supersymmetry which is broken softly.  In addition, however, it has
an exact discrete symmetry known as $R$-parity $R_P$ -- related to baryon
number $B$, lepton number $L$ and spin $S$ via $R_P = (-1)^{3B+L+2S}$ --
under which each SM particle is even while its superpartner is odd.  (At
the superfield level this is the same as matter parity under which quark
and lepton superfields are odd while gauge and Higgs superfields are
even.)  Consequently, superparticles have to be produced in pairs and the
lightest superparticle ($LSP$) is stable and neutral, the latter from
cosmological considerations [3].  On account of its feeble interactions
with ordinary matter, the $LSP$ -- once produced -- escapes detection,
leading to a mismatch in the total measured momentum.  This is the classic
$\ptslash $-signature of superparticle pair-production, the
absence of which so far has led to interesting lower bounds [4] on the
superparticle masses.

The gauge interactions of the MSSM are completely fixed by its particle
content and the gauge group.  The same is not true of its Yukawa terms,
though.  These arise from the $F$-part of a trilinear superpotential and
possess a lot of freedom even after obeying gauge and supersymmetry
invariance.  The additional requirement of $R_P$-conservation restricts
the residual Yukawa terms to be
$$
{\cal L}_Y = \left[h_{ij} L_i H_1 E^C_j
+ h^{\prime}_{ij} Q_i H_1 D^C_j + h^{\prime\prime}_{ij} Q_i H_2
U^C_j\right]_F.
\eqno (1)
$$
In (1) $L$ and $E^C$ ($Q$ and $U^C$, $D^C$) are the lepton doublet and
antilepton singlet (quark doublet and antiquark singlets) left-chiral
superfields, respectively, while $H_1$ ($H_2$) is the Higgs doublet
superfield with weak hypercharge $Y = -1(+1)$.  Moreover, $i$ and $j$ are
generation indices while $h$, $h^\prime$ and $h^{\prime\prime}$ are
coupling strengths.

The possibility of other viable alternatives to the MSSM (violating $R_P$
and leading to very different phenomenology since the classic $\ptslash
$ signature is vitiated by the unstable nature of the $LSP$) has led
several authors [5,6] to study the observable consequences of $R$-parity
breaking models.  Unlike the SM, supersymmetric models do allow for the
possibility of $B$- and $L$-(and hence $R_P$-) violating interactions
which have the most general form:
$$
{\cal L}_{\not R_P } =
\left[\lambda_{ijk} L_i L_j E^C_k + \lambda'_{ijk} L_i Q_j D^C_k +
\lambda^{\prime\prime}_{ijk} U^C_i D^C_j D^C_k\right]_F,
\eqno (2)
$$
where we have used field redefinitions to rotate away bilinears of the
form $L_iH_2$.  The coupling constant matrices $\lambda
{}~(\lambda^{\prime\prime})$ are antisymmetric in the first (last) two
indices.  The first two terms in (2) lead to $L$-violation whereas the
last one causes baryon non-conservation.  The simultaneous presence of both

$B$- and $L$-violating operators would, however, lead to an amplitude for
proton decay suppressed only by $1/m^2_{\tilde q} \lsim 1/(1~{\rm
TeV})^2$.  Thus, at most, only one of these classes of operators can
exist.  For instance, one could have $L$-conservation and $B$-violation,
i.e. $\lambda_{ijk} = 0 =
\lambda^\prime_{ijk}$ and $\lambda^{\prime\prime}_{ijk} \not= 0$ in
general.  There have been [7] cosmological arguments implying strong upper
limits $(\lambda^{\prime\prime} < 10^{-7})$ on
$\lambda^{\prime\prime}_{ijk}$ from the requirement that GUT scale
baryogenesis does not get washed out, though recent studies [8] suggest
that these arguments are model-dependent.  More important for our purpose
is the kind of experimental signals that these different interactions lead
to. As has been shown [9], it may be very difficult to discern signals of
$B$-violating interactions (especially at hadron colliders) above QCD
backgrounds.

These considerations lead one to consider the alternative scenario [8-12]
for $R_P$-breaking, namely $B$-conservation and $L$-violation (i.e.
$\lambda^{\prime\prime}_{ijk} = 0$ and $\lambda_{ijk} \not= 0 \not=
\lambda^\prime_{ijk}$ in general).  If $\lambda^{\prime\prime}_{ijk} = 0$,
the $\lambda^\prime$-terms would need some other baryon-number violating
but $B$-$L$ conserving process, such as non-perturbative instanton-induced

electroweak baryon non-conservation, to wash out the GUT-generated baryon

asymmetry of the universe.  The latter interaction, however, conserves
${1\over3}B$-$L_i$ where $L_i$ is the family lepton number for each lepton
family of type $i$.  Thus the effective conservation of any one lepton
generation would suffice [10] for the retention of the initial baryon
asymmetry so that the cosmological constraints can be satisfied if the
smallest lepton non-conserving Yukawa coupling (where no third generation

lepton need be involved) is less than $10^{-7}$.  This then leaves largely
untouched the strongest possible such coupling (involving a single third
generation lepton) which can now be safely speculated to be $\gsim
10^{-5}$ leading to quite characteristic signals, as discussed below.

As is clear from the previous discussion, all $R_P$-violating models must
necessarily treat quarks and leptons differently, (vis-a-vis their
conserved quantum numbers), in order to be compatible with the absence of
rapid proton decay.  This may appear somewhat contrary to the grand
unification philosophy which tries to put quarks and leptons on a similar
footing.  However, Hall and Suzuki [5] have constructed a grand unified
model in which $R_P$ is violated in the low energy superpotential only by
bilinear terms of the form $L_iH_2$ which can be rotated away by a field
redefinition.  $R_P$-violation then shows up in the lepton non-conserving

trilinear operators, rather than in the baryon non-conserving ones.  In

Unified String Theories, also, there arise [13] discrete symmetries which
treat baryons and leptons differently.  In particular, it has recently
been shown [14] that, consistent with the particle content of the MSSM and
the observed lack of fast proton decay, two such discrete symmetries are
possible: $R_P$ and $B$.  Whereas the former directly eliminates only the
dimension-four contributions to the proton-decay amplitude and not the
dimension-five ones, the latter removes both.  Thus we find it not
unreasonable to work in a $B$-conserving, $R_P$- and $L$-violating
scenario.

There exist various strong upper limits [6] on several of the $\lambda$
and $\lambda^\prime$ terms, as discussed below in Section 2.
Nevertheless, some of these coupling strengths can be $O(10^{-1})$.  As we
will see in Section 2, the lepton-non-conserving trilinear operator with a

single third-generation lepton superfield is
relatively unconstrained and we
shall take that to be the dominant term in the superpotential.  Thus our
assumption is that only $\tau$-number (and not $e$- or $\mu$-number) gets
violated.  We are motivated to consider this since neither neutrinoless
nuclear double-beta decay nor the production of positive muons in nuclear
$\mu$-capture has been observed.  These put stringent restrictions on any
violation of $e$- or $\mu$-number whereas such restrictions are absent for
$\tau$-number.  All constraints from the observed lack of flavour-changing

tau decays can be met with the assumptions of $e$- and $\mu$-conservation
leaving the scope to violate $\tau$-number with impunity.  {\it The prospect of

detecting $\tau$-number violating interactions via the production of $LSP$
pairs produced in $e^+e^-$  annihilation at LEP 200 forms the subject of this
paper.}

In our analysis we shall assume that the $R_P$-violating coupling is large
enough for the $LSP$ to decay inside the detector.  This is ensured
[7,9] by the use of Dawson's [5] calculation of the $LSP$ lifetime and

requiring
$$
\lambda,\lambda^\prime \gsim 5 \times 10^{-7} (m_{\tilde\ell,\tilde
q}/100~{\rm GeV})^2 (100~{\rm GeV}/m_{LSP})^{5/2}.
\eqno (3)
$$
For values of $\lambda~(\lambda')$ that violate the lower bound (3), the
$LSP$ escapes detection so that the signals for superparticle production
at $LEP$ would be essentially the same as their MSSM counterparts [15].
For an $LSP$ mass-range of 20-100 GeV, of interest to us, values of
$\lambda$ and or $\lambda' \gsim 10^{-5}$ would be sufficient to observe
its decay in the apparatus.  The pair-production and subsequent decays of
the $LSP$s will be signalled by the presence of distinctive tau
signatures.  The identification of large $p_T $  $\tau'$s through their
hadronic decay products[16] -- specific mesons such as $\pi, \rho, A_1 $
etc. as well as low-multiplicity
narrow jets -- encourages us to believe that it
will be possible to identify the hadronic decays of the $\tau$ in the
cleaner environment of LEP 200 with  reasonable efficiency.

With $R_P$ not conserved, the cosmological constraints [3] -- requiring
the $LSP$ to be colour and electrically neutral -- no longer apply.  A
priori, the $LSP$ could now be any superparticle.  The
squark (apart from $\tilde t$), however, is an unlikely $LSP$ candidate.
This may be seen as
follows.  If the running squark mass $m_{\tilde q}$ at low energies
is much smaller than the corresponding gluino mass $m_{\tilde g}$,
renormalization group evolution drives $m^2_{\tilde q}$ to negative values
below the unification scale [17] -- leading to colour- and charge-breaking
vacua -- unless large Yukawa interactions are present.  Since the Yukawa
couplings of all but $t$-squarks are generally negligible (and we
exclude the exceptional case [18] of large bottom Yukawa interactions for
$\tan \beta \simeq m_t/m_b$), we can assume that, among squarks, only the
$\tilde t$ could be the $LSP$.
Indeed, the lower $\tilde t$ mass eigenstate [19] may well become
lighter than other superparticles by virtue of $\tilde t_L - \tilde t_R$
mixing induced by soft supersymmetry breaking $A$-terms.

In models with a common gaugino mass at the unification scale, the gluino

is heavier than the $SU(2)$ and $U(1)$ gauginos [1,20], and hence can be

excluded from the $LSP$ list.  This leaves us with the charged sleptons,
the sneutrinos, the charginos and the neutralinos as candidates for the
$LSP$.  However, $LEP$ searches require the masses of charginos, charged
sleptons and sneutrinos to essentially exceed $M_Z/2$, so that the
lightest neutralino $\tilde Z_1$ is really the only candidate for an $LSP$
lighter than 45 GeV.  In order to be definite, we will assume for the most
part that the $LSP$ is indeed a neutralino, though we will qualitatively
discuss how signals are altered in the various other cases.

The cross section for pair-production at LEP 200 is fixed by gauge
interactions and hence is the same as in the MSSM.  Each $LSP$, thus
produced, decays within the apparatus either leptonically by a
$\lambda$-term or semileptonically by a $\lambda^\prime$-term.  There will
be spectacular observable multilepton-final state configurations in the
former case with essentially no background from the SM or any other
non-standard scenario e.g.  $\tau\tau\bar e\bar e\  + \etslash$,

$\bar\tau\bar\tau e e\  + \etslash$, $e\bar e e\bar\tau\  + \etslash$,
$e\bar e\bar e\tau\  + \etslash$ and $e\bar e e\bar e\  + \etslash$.
Additionally, there should be signals for final-state configurations such
as $e\bar e\tau\bar\tau\  +$ $\etslash$ and $\tau\bar\tau
\tau\bar\tau\  + \etslash$ where the backgrounds may be more problematic.
Turning to the $\lambda^\prime$-case, some characteristic observable final
state configurations are $\tau\tau(4j)$ and $\bar\tau\bar\tau (4j)$
whereas one will also have more background-ridden configurations such as
$\tau\bar\tau(4j)$, $\tau(4j)\  + \etslash$, $\bar\tau(4j)\  + \etslash$
etc.  It may be noted that $\tau\tau (4j)$ and $\bar\tau\bar\tau (4j)$
events without $\etslash$ will provide unambiguous evidence for
$\tau$-number non-conservation.  The bulk of our work is devoted to a

discussion of many novel signatures for these processes in explicit $R_P$-
and $L_\tau$-violating models for various possible LSP candidates. We also
highlight interesting interrelations among the different cross sections.

There is a somewhat different version [21] of the $R_P$- and $L$-violating
scheme in which these discrete symmetries suffer spontaneous
breakdown.  However, this scenario cannot obtain within the minimal
particle content of the MSSM (e.g. any $VEV$ attributed to one of the SM
sneutrinos leads to one or more additional decay channel for the $Z$ in
conflict with experiment [22]).  An additional $SU(2)_L \times U(1)_Y$
singlet left-chiral neutral lepton superfield $N$ is required and a VEV
needs to be attributed to its scalar component.  Though this model does
not engage our main concern, we do mention it briefly.  We will also study
rival new physics mechanisms which can mimic our signals, (e.g. a heavy
Majorana neutrino) and discuss how these can be distinguished
from $R_P$-violating processes.

The rest of the paper is organized as follows.  Section 2 contains the
basic $R_P$-violating vertices and interactions.  In Section 3 all our
proposed $R_P$- and $L_\tau$-violating processes, together with their
signatures at LEP 200, are discussed for the case when the $LSP$ is a
neutralino as well as for the other $LSP$ candidates.  We include a
quantitative discussion of the cross sections for $R_P$-violating signals
from neutralino $LSP$s in Section 4.  In Section 5 we discuss some rival
new physics mechanisms which can mimic our signals and suggest ways of
discriminating between them.  Finally, Section 6 contains a summary and
discussion of our results.  The Appendix includes an explicit model of an
unstable heavy Majorana neutrino which is largely an $SU(2)_L$ doublet.
\bigskip

\centerline{\fourteenbf 2. $\rpslash $ vertices and interactions}

\nobreak
We begin by writing the Lagrangian density for ${\rpslash }$
interactions.  The $\lambda$-terms in the superpotential (2) lead to [6],
$$
{\cal L}_{\not R_P,\lambda} = \lambda_{ijk} \left[{\tilde\nu}_{iL} \bar
e_{kR} e_{jL} + {\tilde e}_{jL} \overline{e_{kR}} \nu_{iL} + {\tilde
e}^\star_{kR} \overline{(\nu_{iL})^C} e_{jL} - (i \leftrightarrow
j)\right]_F + h.c.,
\eqno (4)
$$
whereas the $\lambda'$ terms yield
$$
\eqalignno{
{\cal L}_{\not R_P,\lambda'} = \lambda'_{ijk} \Bigg[{\tilde \nu}_{iL}
\overline{d_{kR}} &d_{jL} + {\tilde d}_{jL} \overline{d_{kR}} \nu_{iL} +
{\tilde d}^\star_{kR} \overline{(\nu_{iL})^C} d_{jL} - {\tilde e}_{iL}
\overline{d_{kR}} u_{jL} & \cr & - {\tilde u}_{jL} \overline{d_{kR}}
e_{iL} -
{\tilde d}^\star_{kR} \overline{(e_{iL})^C} u_{jL}\Bigg] + h.c. & (5)}
$$
In (4) and (5) particle names are used to label the corresponding
particle fields.

Many of the couplings in (4) and (5) are already restricted by
experimental data.  For definiteness, we will consider here only those
interactions for which $1\sigma$ constraints from experiments allow the
corresponding $\lambda$ or $\lambda'$ to exceed 0.2 (assuming a sfermion
mass of 200 GeV).  This should be compared with the electromagnetic
coupling $e \simeq 0.3$.  We then see from Table 1 of Ref. 6 that for the
purely leptonic interactions in (4), only the couplings $\lambda_{131}$
and $\lambda_{133}$ satisfy this requirement.  In contrast, the analysis
of Ref. 6 does not lead to any constraint on the couplings,
$\lambda'_{3jk}$ (for all $j$ and $k$), $\lambda'_{222}$, $\lambda'_{223}$,
$\lambda'_{232}$ and $\lambda'_{233}$; furthermore, the couplings
$\lambda'_{121}$, $\lambda'_{122}$, $\lambda'_{133}$, $\lambda'_{123}$ and
$\lambda'_{131}$ can indeed be larger than 0.2, and so satisfy our
requirement above.

As noted in Ref. 23, the experimental upper limit on the mass of the
electron neutrino translates into the bound $\lambda_{133} \lsim 3
\times 10^{-3} \left(m_{\tilde\tau}/100~{\rm GeV}\right)^{1/2}$.
Since we generalize this result, let us recapitulate the argument leading
to it.  We begin by noting that the $\lambda_{133}$ interaction can induce
a Majorana mass,
$$
\delta m_{\nu_e} \sim {\lambda_{133}^2 \over 8\pi^2} {1 \over
m^2_{\tilde\tau}} M_{SUSY} m^2_\tau,
\eqno (6)
$$
for $\nu_e$ via diagrams involving $\tau\tilde\tau$ loops.  In (6), one
factor of $m_\tau$ arises from the $\tau$-chirality flip whereas a factor
$m_\tau~M_{SUSY}$ comes from $\tilde\tau_L - \tilde\tau_R$ mixing.  Taking
$M_{SUSY} \simeq m_{\tilde\tau}$  leads to the bound $\lambda_{133} \lsim
O(10^{-3})$ mentioned above.  It should be clear that the
argument also carries over for the $\lambda'_{1jk}$ couplings in (5); we
then find
$$
\delta m_{\nu_e} \sim {\lambda^{\prime 2}_{1jk} \over 8\pi^2} {1 \over
m^2_{\tilde q}} M_{SUSY} m_j m_k,
\eqno (7)
$$
where $m_j$ and $m_k$ are the masses of the $T_3 = -1/2$ quarks of the
$j^{\rm th}$ and $k^{\rm th}$ generation.  Assuming $m_{\hat q} \simeq
m_{\tilde\tau}$, we
see that the bound on $\lambda'_{1jk}$ is weakened from that on
$\lambda_{133}$ by a factor $\left(m_j m_k/m^2_\tau\right)^{1/2}$. Thus we
derive the hitherto unnoticed constraint that
with the exception of $\lambda'_{112}$,
$\lambda'_{121}$ and $\lambda'_{111}$, the bound on $m_{\nu_e}$ excludes
all $\lambda'_{1jk}$ type couplings.

Combining the results of this analysis with those of Ref. 6 discussed
above, we see that the allowed couplings are just $\lambda_{131}$,
$\lambda'_{3jk}$, $\lambda'_{121}$, $\lambda'_{222}$, $\lambda'_{223}$,
$\lambda'_{232}$ and $\lambda'_{233}$.  The first two couplings violate
only $\tau$-number conservation, the third violates $e$-number
conservation, while the remaining ones violate the conservation of muon
number.  We should also mention that there may be further constraints on
the simultaneous violation of two, or more, lepton flavours, since then
restrictions from the non-observation of $\mu \rightarrow e$, $\tau
\rightarrow \mu$ transitions will also be applicable.

In the following, we will mainly focus on the possibility that just
$\tau$-lepton number is violated.  The relevant vertices are shown in Fig.
1 and Fig. 2 for the $\lambda$ and $\lambda'$ type interactions,
respectively.  It will be easy to adapt our discussion for the case where,
instead, $e-$ or $\mu$-violating interactions are dominant.  Since the

efficiency for the identification of $\tau$'s is significantly smaller
than that for $e$ or $\mu$, we expect that it will be considerably easier
to identify signals for $e$- or $\mu$-violating interactions.  We will
return to these issues in the concluding section.
\bigskip

\centerline{\fourteenbf 3. $\rpslash$ and $\not L_\tau$ processes}

\nobreak
Our prototype process consists of the simultaneous decay of a pair of
$LSP$s (assumed to be the neutralino $\tilde Z_1$) produced at LEP 200.
The $R_P$-conserving production reaction $e^+e^- \rightarrow \tilde Z_1
\tilde Z_1$ gets followed by each $\tilde Z_1$ undergoing an
$R_P$-violating decay into three fermions -- changing $L_\tau$ by one
unit.  The three-body decay of each $\tilde Z_1$ proceeds first by a gauge
vertex transition into a real fermion and a virtual sfermion, the latter
further undergoing a transition into two additional fermions via one of
the vertices of Fig. 1 or Fig. 2.  If the virtual sfermion is a
third-generation slepton, the $R_P$- and $L_\tau$-violating vertex
can be either
of the $\lambda$-type or of the $\lambda'$-type.  In the former case the
decay products of the $\tilde Z_1$ are two oppositely charged leptons
which are visible and a neutrino which generates $\etslash$, i.e. $\ell
\bar \ell'\  + \etslash$.  In the latter case the decay products are a
$\tau$ or a $\nu_\tau$, accompanied by two quarks which generally fragment
into two jets, i.e. $(2j)\ell$, $(2j)\bar \ell$
 or $(2j)\  + \etslash$.  Here and in the following we denote each
quark as an independent jet, though the jets could actually merge.  For
the situation where the virtual sfermion is a squark, the corresponding
$R_P$-violating vertex must necessarily be of the $\lambda'$-type so that
the decay products of the $\tilde Z_1$ appear as one of three possible
combinations: $(2j)\tau$, $(2j)\bar\tau$ and $(2j)\  + \etslash$.

The pair of on-shell $LSP$s, produced in $e^+e^-$ annihilation, can
finally lead to three types of visible final state configurations
corresponding to three possible combinations of $\lambda$- and
$\lambda'$-type decay vertices involved in the transition of the virtual
sfermion.

(1) Both operative $R_P$- and $L_\tau$-violating vertices are of the
$\lambda$-type resulting in four charged leptons (of total charge zero)
and $\etslash$ from the two $\tilde Z_1$'s.

(2) The decays of both $\tilde Z_1$'s involve $\lambda'$-type vertices
yielding one of the following six visible final state configurations:
$(4j) \tau \tau, (4j)\tau\bar \tau, (4j)\bar\tau \bar \tau, (4j)\tau\
+ \etslash, (4j)\bar \tau\  + $ $\etslash$ and $(4j)\  + \etslash$.

(3) One $\tilde Z_1$-decay involves a $\lambda$-type vertex and the other
a $\lambda'$-type interaction leading to the following nine visible
combinations: $\tau\tau\bar e(2j)\  + \etslash$, $\bar\tau\bar\tau
e(2j)\  + \etslash$, $\tau\bar\tau e(2j)\  + \etslash$, $\tau\bar\tau\bar
e(2j)\  + \etslash$, $e\bar e\tau(2j)\  + \etslash$, $e\bar e\bar\tau
(2j)\  + \etslash$, $\tau\bar e(2j)\  + \etslash$,
$\bar\tau e(2j)\  + \etslash$
 and $e\bar e(2j)\  + \etslash$.

Let us take case (1) above first.  As discussed in Section 2, only the
coupling $\lambda_{131}$ is allowed.  The possible decay products of the
$\tilde Z_1$ from a $\lambda_{131}$ vertex are $\tau\bar e\nu_e$, $\bar
\tau e\bar \nu_e$, $\bar e e\nu_\tau$ and $e\bar e\bar \nu_\tau$.  At the
tree level each decay proceeds via three diagrams separately involving
stau-exchange,
selectron-exchange and sneutrino-exchange.  In the limit of ignoring the
masses of all final state leptons and of taking all sfermions to be
mass-degenerate, all the partial widths are identical.  In what follows,
we give ratios of cross sections rather than observable rates which have
to be calculated by folding in the appropriate detection efficiencies.
The total cross sections for the six visible distinct final state
configurations formed out of $e^+e^-$ collision will be in the
combinatorial ratios $$
\eqalign{
\sigma(\tau \tau\bar e\bar e\  +~\etslash)&:\sigma(\bar\tau\bar\tau ee\
+~\etslash):\sigma(e\bar e\tau\bar \tau\  +~\etslash):\sigma(e\bar e\bar
e\tau\  +~\etslash):\sigma(e\bar e e\bar\tau\  +~\etslash)\cr
&:\sigma(e\bar
ee\bar e\  +~\etslash) = 1:1:2:4:4:4.}
\eqno (8)
$$
Rate estimates will be provided in Section 4. We just comment
here on the fact that in the $SM$ the processes $e\bar e \rightarrow
\tau\tau \bar e\bar e\  + \etslash$,
$\bar \tau\bar \tau e e\  + \etslash $
, $e\bar e
\rightarrow e\bar e\tau\bar \tau\  + \etslash$,
$e\bar e e\bar \tau\  + \etslash$
, $e\bar e\bar e\tau\  + \etslash$ and $e\bar e e\bar e\  + \etslash$
have rather tiny rates.

Turning to case (2), the possible decay products of the $\tilde Z_1$ from
all $\lambda'_{3jk}$ vertices are $\tau u_j \bar d_k$, $\bar \tau \bar u_j
d_k$, $\nu_\tau d_j \bar d_k$ and $\bar \nu_\tau \bar d_j d_k$.
Since the top quark is kinematically inaccessible in LSP decays, the
generation index $j$ runs over just 1,2 for up-type quarks and 1,2,3 for
down-type quarks.
Moreover, each decay proceeds via three tree diagrams --
exchanging a left slepton, a left squark or a right squark.  Working in
the same mass limit
mentioned earlier, the combinatorial rate ratios between the five visible

final state configurations now are:
$$
\eqalign{
\sigma[\tau\tau(4j)]&:\sigma[\bar\tau\bar\tau (4j)]:\sigma[\tau\bar\tau
(4j)]:\sigma[\tau (4j)\  +~\etslash]:\sigma[\bar\tau(4j)\  +~\etslash]\cr
 & :\sigma[(4j)\  +~\etslash] = 1:1:2:2x:2x:x^2.}
\eqno (9)
$$
The factor $x$ arises from the fact that the top quark is not produced.
It is equal to $2 + |\alpha|^2$, $\alpha$ being a coupling-dependent

parameter which vanishes if $\lambda'_{33k} = 0$, and diverges if
$\lambda'_{33k}$ are the dominant couplings.
The reactions $e^+e^- \rightarrow \tau\tau(4j)$, $\bar\tau\bar\tau (4j)$
are specially interesting in that there is no missing $E_T$ in the primary
process.  The possibility of searching for $\tau-$number violation via
like sign ditau signal, first proposed for hadron colldiers in Ref. 2,
holds even better promise at LEP 200.
These reactions are
hallmarks of the self-conjugate nature of the $LSP$s and would be
essentially absent in the $SM$.  A $q\bar q$ pair and two radiated gluons

plus a virtual photon decaying into a $\tau$-pair could yield $\tau\bar\tau
(4j)$ but with a tiny rate.  The $\tau(4j)\  + \etslash$ or $\bar \tau
(4j)\  + \etslash$ final state could come from two $W$'s, one decaying
semileptonically into a $\tau$ (or $\bar \tau$) and the other into $q\bar
q'$ plus two radiated gluons; but the rate would again be rather low.  The
$(4j)\  + \etslash$ final state could arise from double
$Z$ production, one
$Z$ decaying into a $\nu\bar \nu$ pair and the other into $q\bar q$ plus
two radiated gluons.

Lastly, in case (3), arguments -- similar to those given above and in the
same limit -- imply:
$$
\eqalign{&
\sigma[\tau\tau\bar e(2j)\  +~\etslash]:\sigma[\bar\tau\bar\tau
e(2j)\  +~\etslash]:\sigma[\tau\bar\tau e(2j)\  +~\etslash
:\sigma[\tau\bar\tau\bar e(2j)\  +~\etslash]\cr & ~~~~~:\sigma[\bar
ee\tau(2j)\  +~\etslash]:\sigma[\bar \tau e\bar e(2j)\  +~\etslash
:\sigma[\tau\bar e(2j)\  +~\etslash]:\sigma[\bar\tau
e(2j)\  +~\etslash]\cr & ~~~~~:\sigma [\bar ee(2j)\  +~\etslash] =
1:1:1:1:2:2:2x:2x:4x.}
\eqno (10)
$$
Except for the last configuration, all the others are quite striking and
difficult to simulate in the $SM$.

We will now discuss, within our explicit $R_P$- and $L_\tau$-breaking
scenario, the consequences of the $LSP$ being different from a neutralino.
As explained in the Introduction, there are theoretical reasons that
disfavour squarks (except, possibly, a light stop) and gluinos from being

$LSP$ candidates so that after the lightest neutralino  we need consider
only the lightest electroweak chargino $\tilde W_1$, the lightest slepton
and the strongly interacting stop $\tilde t$.  In any case, LEP
experiments have [4,24] established a lower mass bound in the vicinity of
$M_Z/2$.  However, the magnitudes of the  cross sections concerned are not
very sensitive to the mass of the $LSP$ unless it is at the boundary of
phase space.  Moreover, in the mass-range of interest, the cross section
for chargino pair-production is substantially larger than that for
neutralinos, while slepton or stop particle-antiparticle pairs would be
produced at smaller rates.

Turning to event characteristics, consider the chargino case first.
Exactly as in the neutralino case, it can decay into a fermion (quark or

lepton) antifermion pair in which one is on-shell and the other is
off-shell.  The latter, if a lepton, decays only by a $\lambda$-type
coupling while, if a quark, it can decay either by a $\lambda$-or by a
$\lambda'$-term.  Once again there are three possibilities:

1) The decay of each chargino $\tilde W_1$ proceeds via the $\lambda_{131}$
coupling.  There are two channels, $\tilde W^-_1 \rightarrow e\bar e\tau$

and $\tilde W^-_1 \rightarrow e\bar\nu_e\bar\nu_\tau$, as well as their

charge conjugates.  Each has two tree-level diagrams mediated by a first-

or third-generation virtual slepton, down-type for the first channel and
up-type for the second.  In the limit specified earlier, the corresponding
partial decay widths are in the ratio $\sin^2 \gamma_L:\sin^2\gamma_R$.
Here we use the notation of Baer et. al. [25] with $\gamma_{L,R}$ as the
rotation angles in the mass-diagonalization of the left-, right-handed
wino fields.  The total rates for the four visible final state
configurations in $e^+e^-$ annihilation will be in the ratio
$$
\eqalign{
\sigma(e\bar ee\bar e\tau\bar \tau)&
:\sigma(e\bar e\tau\bar e\  +~\etslash
):\sigma(e\bar e\bar\tau e\  +~\etslash):\sigma(e\bar e\  +~\etslash)
= \cr &
\sin^4\gamma_L:\sin^2\gamma_L\sin^2\gamma_R:\sin^2\gamma_L\sin^2\gamma_R:
\sin^4\gamma_R.}
\eqno (11)
$$

2) Each $\tilde W_1$ decays by use of a $\lambda'$-vertex.  The decay
channels are $\tilde W^-_1 \rightarrow \tau d_j\bar d_k$ and $\tilde W^-_1
\rightarrow \bar \nu_\tau d_j \bar u_k$ as well as their charge
conjugates.  Once again, there are two tree-level diagrams per channel
involving squark and slepton exchanges: down-type for the first channel and
up-type for the other.  Now, because of the absence of the top from the
final state, the partial widths are
as $y$ $\sin^2\gamma_L:\sin^2\gamma_R$ where $y$ has the form $1 +
|\beta|^2$, $\beta$ being a parameter analogous to $\alpha$.  The total
cross sections of the four visible final-state configurations are expected
to be produced in the ratios:
$$
\eqalign{
\sigma[\tau\bar\tau(4j)]&:\sigma[\tau(4j)\  +~\etslash
]:\sigma[\bar\tau(4j)\  +~\etslash]:\sigma[(4j)\  +~\etslash] \cr &
= y^2\sin^4\gamma_L:y\sin^2\gamma_L\sin^2\gamma_R:
y\sin^2\gamma_L\sin^2\gamma_R
:\sin^4\gamma_R,}
\eqno (12)
$$

3) One $\tilde W_1$ decays via a $\lambda$-type vertex and the other
through a $\lambda'$-type one.  There are seven different visible final
state configurations now with total rate proportionalities given by
$$
\eqalign{
\sigma[e\bar e\tau\bar\tau(2j)]&:\sigma[e\bar e\tau(2j)\  +~\etslash]
:\sigma[e\bar e\bar\tau (2j)\  +~\etslash]:\sigma[e\bar \tau
(2j)\  +~\etslash] \cr & :\sigma[\bar e\tau(2j)\  +~\etslash]
:\sigma[e(2j)\  +~\etslash]:\sigma[\bar e(2j)\  +~\etslash] \cr &
= \eqalign{&
2y\sin^4\gamma_L:\sin^2\gamma_L\sin^2\gamma_R:\sin^2\gamma_L\sin^2
\gamma_R:y
\sin^2\gamma_L\sin^2\gamma_R \cr &
:y\sin^2\gamma_L\sin^2\gamma_R:\sin^4\gamma_R:\sin^4\gamma_R.}}
\eqno (13)
$$
Evidently, there are many striking event configurations here which would
be hard to produce in the $SM$.  But, because of the Dirac nature of the
chargino, there are no unambiguous indicators
of $\tau$-number violation.

Turning now to sleptons, in the case where $\tilde\nu_\tau$ is the $LSP$,
it can be
produced in $e^+e^-$ collision either singly via the $\lambda_{131}$
coupling (Fig. 1) or in a pair through gauge interactions.  The decay of a
$\tilde\nu_\tau$ can take place either into $e\bar e$ or into $\bar d_j
d_k$ by means of the $\lambda'_{3jk}$
couplings (Fig. 2).  The presence of an $s$-channel resonance will be a
spectacular indicator of the former.  However, being of small width, it
may easily be missed at LEP 200 unless there is a dedicated search spanning
the CM energy range $100-200$ GeV in narrow bins.  On the other hand, if
$|\lambda_{131}|$ is much less than the semiweak gauge coupling strength,
pair-production would really be the dominant mechanism to produce
$\tilde\nu_\tau$'s in $e^+e^-$ collision.  Considering only the latter
process, the different possible visible final-state configurations will be
$e^+e^-e^+e^-$, $e^+e^-(2j)$ and $4j$; these should lead to an
observable increase in the number of spherical events at LEP 200.  In the
first and
second cases, each of the appropriate $e^+e^-$ pair(s) will have a resonant

invariant mass facilitating a relatively clean separation of these events.
However, we find no clear $L_\tau$-violating signature in the case where

$\tilde\nu_\tau$ is the $LSP$, since any pair-produced scalars

decaying into $e^+e^-$ or $q\bar q$ will generate similar signals.

In the case where the $LSP$ is a sneutrino belonging to either of the
first two
generations, it can only be pair-produced at $e^+e^-$ colliders.  Since we
retain only the couplings $\lambda_{131}, ~\lambda'_{3jk}$ in (6) and (7),
only $\tilde\nu_e$ and $\tilde\nu^\star_e$ can have direct two-body decays
at the tree level (Fig. 1): $\tilde\nu_e \rightarrow e\bar \tau$,
$\tilde\nu^\star_e \rightarrow \bar e\tau$.  In general, two-body decays of
$\tilde\nu_\mu$ and $\tilde\nu^\star_\mu$ are also possible but they
can only take
place through 1-loop diagrams making them longer-lived.  These
decays are $\tilde\nu_\mu \rightarrow
\mu\bar\tau,\nu_\mu\nu_\tau,\nu_\mu\bar\nu_\tau$ and the corresponding
conjugates. They are, however, absent if $\lambda_{131}$ is the only
$R_P$-violating coupling. Thus it is more likely that $\tilde\nu_\mu$
would decay via four-body modes. These decays proceed in three steps, as
shown in Fig 3, resulting in the final states
$\mu\bar\tau f\bar f,\nu_\mu \tau f_1
\bar f_2,\nu_\mu\bar\tau\bar f_1 f_2$ and their conjugates.
Here $f$ is either $e$ or a $d_k$-quark and the $f_1\bar f_2$ pair can be
either $\nu_e\bar e$ or $u_j \bar d_k$.

An interesting point in connection with the four-body decays is the
following.  In each two-body decay the sign of the emanant $e$ or $\mu$
(and hence that of the associated $\tau$) is determined by $e$- or
$\mu$-conservation.  Thus, as in the $LSP = \tilde\nu_\tau$ case, there is
no direct evidence of $\tau$-number violation.  In contrast, in the $\tilde
Z_1$-mediated four-body decays, the emanant $\tau$ from the same decaying
sneutrino can have either sign.  Thus one can have -- in $e^+e^-$
collisions -- eight different visible leptonic final configurations from
$\tilde\nu_\mu \tilde\nu^\star_\mu$ $LSP$ pair-production:
$\mu\bar\mu\tau\bar\tau e\bar e e\bar e$, $\mu\bar\tau\bar\tau e\bar e
e\  +\etslash$, $\bar \mu \tau \tau \bar e e\bar e\  +\etslash$,
$\mu\bar \tau \tau e\bar e\bar e\  +\etslash$,
$\bar\mu \tau \bar \tau\bar e e
e\  +\etslash$, $\tau\bar \tau e\bar e\  +\etslash$,
$\tau\tau\bar e\bar e\  +\etslash$ and
$\bar\tau\bar\tau ee\  +\etslash$.
Additionally, there can be seventeen different visible lepton-jet
combinations:  $\mu\bar \mu\tau\bar\tau (4j)$, $\mu\bar\tau\bar\tau
(4j)\  +\etslash$, $\bar\mu \tau\tau (4j)\  +\etslash$, $\mu\bar
\tau\tau(4j)\  + \etslash$, $\bar\mu \tau\bar \tau(4j)\  + $ $\etslash$,
$\tau\bar\tau (4j)\  + \etslash$, $\tau\tau (4j)\  + \etslash$,
$\bar\tau\bar\tau (4j)\  + \etslash$, $\mu\bar\mu \tau\bar
\tau e\bar e (2j)$, $\mu\bar\tau\bar\tau e\bar e(2j)\  + \etslash$,
$\bar\mu \tau\tau\bar e e(2j)\  + \etslash$,
$\mu\bar\tau\bar\tau\bar e(2j\  + \etslash$,
 $\bar\mu\tau\tau e(2j)\  + \etslash$, $\tau\bar\tau e(2j)\  + \etslash$,
$\bar\tau\tau\bar e(2j)\  + \etslash$,
$\tau\tau\bar e(2j)\  + \etslash$ and
$\bar\tau\bar\tau e(2j)\  + \etslash$.  The cross sections for conjugate
channels are identical.  However, since the branching fractions for the
various decays of $\tilde\nu_\mu$ would depend on the details of the
gaugino-higgsino mixing matrices, we do not make estimates of
the above cross sections in this scenario.  Lastly, note that the analysis
of the signatures for the situation when a charged slepton is the $LSP$
parallels that of the sneutrino case.  Thus we will not elaborate on this
further.

Let us finally consider the case where the $LSP$ is the top squark (or stop)
$\tilde t$.  As we will see, this is rather similar to the $LSP =
\tilde\nu_\ell$ $(\ell \not= \tau)$ case discussed above.  As explained in
Ref. [19], substantial mixing between the $\tilde t_L$ and $\tilde t_R$
states, caused by the large Yukawa interactions of the top family, may
make the lighter of the two stop mass-eigenstates $(\tilde t_1)$ lower in
mass than all other superparticles.  The mass-breaking as well as the
mixing angle between the two $\tilde t$-states depends on various yet
unknown constants such as the supersymmetry-breaking $A$ parameter, the
supersymmetric higgsino mass and $\tan\beta$, the ratio of the two VEVs of
the Higgs fields in the model.  Top squarks can only be pair-produced at
LEP 200.  The $s$-channel photon contribution to the production cross
section is fixed by quantum electrodynamics so that the total production
rate is not expected to be sensitive to the details of stop mixing.
Furthermore, the decay patterns of $\tilde t_1$ are fixed by the
$R_P$-violating interactions (5) so that the qualitative features of the
signals for the production of $\tilde t_1$ pairs are independent of the
unknown details of the $t$-squark sector.  Since we are interested in
signals at LEP 200, we will further assume that $m_{\tilde t_1} < m_t$
since even a $t$-squark as light as 50 GeV can be accommodated, though

other strongly interacting superparticles such as the gluino may have
masses as high as several hundred GeV.

The decays of the lightest $t$-squark will be somewhat similar to those of
the $\tilde\nu_e = LSP$ case discussed earlier.  It can directly decay by
the two-body mode $\tilde t_1 ~\longrightarrow~ \bar\tau d_k$
at the tree-level via the $\lambda'_{33k}$ coupling (Fig. 2).  It can also
have the four-body decay $\tilde t_1 ~\longrightarrow~ \bar\tau b f\bar f$
in analogy with $\tilde\nu_\ell$ (Fig. 4a), only the initial vertex
$\tilde t_1 b\tilde W_1$ being different.  It may be noted, though,
that -- unlike as in the $\tilde\nu_\ell$ case for the LSP -- there will be
no like sign ditau signal from the decays of $\tilde t_1$ and ${\tilde
t}^\star_1$. The sign of the $\tau$ is
determined by that of $\tilde t_1$ since neutralino-mediated $\tau$ decays
would involve a $t$ quark in the final state, which is kinematically
forbidden.  Finally, we remark that, if the
two-body decays dominate over four-body ones, signals from $\tilde t_1$
pair-production will resemble those from the production of $\tau$ scalar
leptoquark pairs.
\bigskip

\centerline{\fourteenbf 4. Cross sections for $\rpslash $ signals from
neutralinos}

\nobreak
In this section we present cross sections for $\tau$-number and
$R_P$-violating signals from the production of neutralinos at LEP 200.
Our reasons for focusing on neutralinos as opposed to other $LSP$
candidates are two-fold. First, in many models, the $LSP$ is likely to be
a neutralino. Second, as we have seen, the Majorana nature of the
neutralino can potentially result in unambiguous signals for $\tau$-number
violation in the form of like-sign ditau events. These will have low
missing $E_T$ coming only from the decay of the taus. The absence of any
large missing $E_T$ thus make it unlikely that there would be two
undetected particles in these events that balance $\tau$-number.

The cross section for the production of a $\tilde Z_1\tilde Z_1$ pair
depends on the mixing angles in the neutralino sector.  Here, we have used
the MSSM as a guide; the cross section $\sigma(\tilde Z\tilde Z_1)$ is
then determined [26] by just a few parameters.  We may take these to be
(1) the gluino mass $(m_{\tilde g})$ which fixes the $SU(2)$ and $U(1)$
gaugino masses via a unification condition [1]; (2) the supersymmetric
higgsino mass, $2m_1$; (3) the ratio, tan $\beta$, of the vacuum
expectation values of the two Higgs fields $H_2$ and $H_1$ and (4) the
selectron mass which enters via amplitudes involving selectron exchange.

Our results for $\sigma(\tilde Z_1\tilde Z_1)$ are shown in Fig. 4 for (a)
$m_{\tilde e} = 100$ GeV and (b) $m_{\tilde e} = 200$ GeV.  We have fixed
tan $\beta = 2$ and illustrated the cross section in the $2m_1 - m_{\tilde
g}$ plane.  The area between the heavy dotted lines corresponds to the
region where $m_{\tilde Z_1} \lsim 45$ GeV.  For parameters in this
region, $\tilde Z_1\tilde Z_1$ pair production should be accessible at LEP
(unless $\tilde Z_1$ is essentially a pure gaugino and the slepton is
heavy).  The decays of the $\tilde Z_1\tilde Z_1$ pair would then lead to
an excess of spherical events including tau leptons.  Although such events
may not have been explicitly searched for at LEP, we should bear in mind
that the absence of such spherical events there can exclude about half the
parameter plane in the scenario that we are considering.

It may be seen from Fig. 4 that, even for $m_{\tilde Z_1} > 45$ GeV,
$\sigma(\tilde Z_1\tilde Z_1)$ may be almost 1 pb provided that $m_{\tilde
e} \simeq 100$ GeV.  We stress that such light sleptons are perfectly
consistent with CDF bounds on squark masses even within the framework of
supergravity models.  Fig. 4b, however, shows that in the ``LEP 200
region'' the cross section falls off rapidly with increasing slepton mass.
This is because over much of this region $|2m_1|$ is rather large so that
the $LSP$ is dominantly a gaugino.  The slepton exchange contribution to
the $\tilde Z_1\tilde Z_1$ production amplitude is then very significant.
Nevertheless, up to a hundred $\tilde Z_1\tilde Z_1$ events are expected

annually even if $m_{\tilde e} \simeq 200$ GeV, assuming an integrated
luminosity of 500 pb$^{-1}$ at LEP 200.  We then see from (8) and (9) that
a handful of like-sign tau events may be expected in this case,
assuming that the taus can be identified by their single prong hadronic
decay which results in isolated, hard $\pi^\pm$ tracks.  If $m_{\tilde e}
\simeq 100$ GeV, the signal may be larger by as much as a factor five.  In
contrast, if the sleptons are very heavy, the signal is likely to be
unobservable.  The dependence of the signal on tan $\beta$ is illustrated
in Fig. 5.  We see that the signal is relatively insensitive to tan
$\beta$ in the region where $m_{\tilde Z_1} \geq 45$ GeV.

We note here that, by combining the ratios (8) - (10) with the results in
Figs. 4 and 5, it is possible to obtain an estimate of the cross sections
for various event topologies from neutralino pair production at LEP 200 if
we assume that either $\lambda$- or $\lambda'$-type operators dominate.
We see that these cross sections are all rather small.  The detectability
of these novel signals, in particular, the like-sign ditau $+$ jets signal,
will
crucially depend on the experimental efficiency for $\tau$-identification.

In order to give the reader some idea of the kinematics of the $LSP$ events,

we have shown in Fig. 6a the $p_T$ distribution of the leptons in the
$\tau\tau \bar e\bar e\  + \etslash$ and
$\bar\tau \bar\tau ee\  + \etslash$
final states that result if the $LSP$
decays by the $\lambda_{131}$ interaction. These have been obtained by
explicitly calculating the concerned matrix element squared with MSSM
couplings [25] and taking all sleptons to be equally massive; for the
slepton masses that we consider, the distributions are essentially governed

by phase space.
We have illustrated these
distributions for $m_{\tilde Z_1} = 45$ GeV and $m_{\tilde Z_1} = 90$ GeV

and, for just the former case, for two values of SUSY parameters which
give rise to different values of $\sigma(\tilde Z_1\tilde Z_1)$.  Also
shown is the $p_T$ distribution from the SM background from $ZZ$
production where both $Z$'s decay via $\tau\bar\tau$, and the electrons
arise via $\tau$-decay.  As expected for the latter, the $p_T(e)$
distribution is very soft.  In contrast, we see that the $p_T$ distribution
of the leptons from $LSP$ decays is fairly hard and essentially determined by

the mass of the $LSP$.  Since electrons with a $p_T$ of a few GeV should
readily be detectable at LEP 200, we believe that the signal will be
determined mainly by the $\tau$ detection efficiency.

Since the $\tau$'s are expected to be identified via their narrow, low-charged
multiplicity $(n = 1 ~{\rm or}~ 3)$ jets, the detectability of
$\tau\tau (4j)$ events will critically depend on how isolated these
$\tau$'s are.  Toward this end, we have constructed a parton-level Monte
Carlo program to simulate these events from neutralino pair production.
Jets are defined to be partons; we have coalesced partons within $\Delta
r \equiv [(\Delta y)^2 + (\Delta \phi)^2]^{1/2} < 0.7$ into a single
jet.  Fig. 6b shows the distribution of

${\rm Min}~\Delta r (\tau,{\rm jet})$ in
the $\tau\tau$ (multi-jet) events, where ${\rm Min}~\Delta r$ is the
minimum separation between either of the $\tau$'s and the nearest jet,
in each event.
In this figure, we have also required that the jets and
$\tau$'s be all central, i.e. satisfy $|y| \leq 1.5$.  We see from the
figure that the $\tau$'s are well separated from the jets.  Even for
$m_{\tilde Z_1} = 30$ GeV, about 2/3 of the events satisfy $\Delta r >
0.5$, whereas for heavy neutralinos this figure is considerably larger.
For instance, if $m_{\tilde Z_1}$ is 60 GeV, the requirement that both the
$\tau$'s satisfy $\Delta r > 0.5$ causes a loss of only 20\% of the events
where all the leptons and jets are central.
We should also mention that the $p_T (\tau)$ distribution in these events
should be similar to that in Fig. 6a. Similar observations apply to the
events containing single $\tau$ ($\bar\tau$) $+$ jets, discussed in
Section 3.

The results of Fig. 6 are encouraging.  We have further checked that the
missing $E_T$ in these events is essentially determined by the $LSP$ mass,
and is typically slightly below $m_{\tilde Z_1}/2$.  Finally, we note that
the $\tau$'s are acollinear.  For $m_{\tilde Z_1} = 30$ GeV, the angular
separation $\Delta\phi$ between the $\tau$'s is, on average,
about $150^\circ$, while
for $m_{\tilde Z_1} \simeq 60$ GeV this becomes $120^\circ$.
We should note, though, that this distribution peaks at
$\Delta\phi=180^\circ$. We also mention that for $m_{\tilde Z_1}\ge 45$
GeV, four jet topologies dominate, whereas for lighter neutralinos, there
is three-jet dominance.
While our
preliminary results appear promising, detailed Monte Carlo studies are
necessary before definite conclusions can be drawn regarding the viability
of these signals.

At this point, several remarks are in order:

\item{(i)} As stated in the Introduction, we focus only on signals
from $LSP$ pair production assuming that all other sparticles are
kinematically inaccessible.  Within the MSSM, charginos will also be
accessible at LEP 200 for a large part of the parameter space in Figs. 5

and 6.

\item{(ii)} We have assumed that the $LSP$ mixing patterns which determine
the cross-sections in Figs. 5 and 6 are as given by the MSSM.  This may,
of course, not be the case so that (in principle) the cross sections may
differ considerably from those shown.  The rates shown in the figures
should only be regarded as indicative.

Before concluding this Section, we note that if the $LSP$ is any sparticle
other than the neutralino, the cross sections may be substantially different

from those shown in Fig. 4 and Fig. 5.  For instance, the cross section
for producing a pair of 60 GeV charginos is [15] typically a few
picobarns, whereas the
corresponding cross section for the case when the $LSP$ is a 60 GeV slepton
or top-squark is about an order of magnitude less.
The latter process also suffers a $p$-wave
suppression so that its cross section falls rapidly with an increasing
sfermion mass.  Finally, we note that if the neutralinos are heavy, the
selectron pair production cross section becomes comparable to that for
smuons or staus; $t$-channel neutralino exchange contributions to
$\sigma_{\tilde e\tilde e}$ may, however, enhance this if $m_{\tilde Z_1}
\approx m_{\tilde e}$.
\bigskip

\centerline{\fourteenbf 5. Alternative mechanisms}

\nobreak
There could be rival ``new physics'' mechanisms that can mimic the $LSP$
signals discussed in Section 3.  For definiteness, let us focus on the
$\tilde Z_1 = LSP$ case.  The distinctive like-sign ditau signals are due
to the Majorana nature of the neutralino.  However, an unstable heavy
Majorana neutrino, containing an admixture of $\nu_\tau$, can also yield
similar $\tau$-number violating signals.

If the heavy Majorana neutrino $\nu_M$ is dominantly an $SU(2)_L$ singlet
with a small component of the usual doublet, the $GIM$ mechanism is no
longer operative.  The decay $Z \rightarrow \nu_M (\nu_\tau)_{\rm phys}$
should then proceed at a rate which is suppressed relative to that of $Z
\rightarrow \nu_e\bar \nu_e$ by a factor of $\sin^2\alpha$, where
$\sin\alpha$ is the $\nu_\tau$-admixture in $\nu_M$.  The subsequent decay
of $\nu_M$ would then lead to spectacular missing-$E_T$ events at LEP.
The non-observation of such events in the sample of $O(10^6)Z$ bosons,
already collected by LEP experiments, then requires that $\sin\alpha \lsim
O(10^{-2})$.  In this case the cross section for the production of a
$\nu_M$-pair, which is suppressed by $\sin^4\alpha$, is too small to be
interesting.

We are thus led to examine the possibility that $\nu_M$ contains a
substantial $SU(2)_L$ doublet component [27] and decays via $\tau$-number
violating interactions.  A simple model realizing this possibility is
presented in the Appendix.  The Majorana neutrino here is essentially a
sequential fourth generation neutrino which gets a mass in the range $50 -
100$ GeV by seesaw mixing with an $SU(2)_L$ singlet neutrino with a
Majorana mass $\sim 1$ TeV.  In this scenario the strength of the $Z\nu_M
\nu_M$ coupling is comparable to that of the $Z\nu_e\bar\nu_e$ one. Thus
the cross section for producing a $\nu_M$-pair may well exceed that for
the pair-production of neutralinos (which is often reduced by mixing angle
factors), shown in Figs. 4 and 5.  It is, therefore, necessary to study
the details of the final states obtained via $\nu_M\nu_M$ production in
order to see if the $\nu_M$ signals can be distinguished from those for
neutralinos.

Once produced, a $\nu_M$ can decay only via gauge
interactions into $\tau\bar\ell\nu_\ell$,
$\bar\tau\ell\bar\nu_\ell$, $\tau u_i\bar d_j$ and $\bar\tau\bar u_i d_j$
via $W$-exchange and into $\nu_\tau\ell\bar\ell$,
$\bar\nu_\tau\ell\bar\ell$, $\nu_\tau q\bar q$ and $\bar\nu_\tau q\bar q$
via $Z$-exchange.  Furthermore, since gauge interactions are universal,
all flavours of quarks and leptons are produced via these decays provided

they are kinematically accessible.

Interesting final state leptonic configurations from $\nu_M\nu_M$
production, therefore, include $\tau\tau\bar\ell\bar\ell'\  + \etslash$,
$\bar\tau \bar\tau \ell \ell'\  + \etslash$,
$\ell \bar\ell' \tau\bar\tau\ + \etslash$, $\tau
\ell\bar\ell'\bar\ell^{\prime\prime}\  + \etslash$ and
$\bar\tau\bar\ell \ell' \ell^{\prime\prime}\   + \etslash$, where
$\ell,\ell'$ and $\ell^{\prime\prime}$ can be $e,\mu$ or $\tau$ with equal
likelihood.  Thus characteristic final states with muons and also with
more than two $\tau$'s, which were essentially absent in the neutralino
case discussed earlier, will also be present.  In addition, since the
charged and neutral current interactions -- involved in the decay of
$\nu_M$ -- are different, the five cross sections in (8) will no longer be
in the specified proportionality.  Turning to the semihadronic decays of
$\nu_M$, we see that the final states are more or less the same as for the
neutralino.  Moreover, the characteristic ditau $+ (4j)$ events are
produced in the same ratio as in (9).  However, for the other channels
described in (9), the full proportionality does not hold because of the
simultaneous presence of $W$- and $Z$-exchange contributions here.  A
similar statement holds also for (10).  There will also be additional
relations among some of these configurations if they are generated from
the decays of a pair of $\nu_M$'s: e.g. for $W$-exchange decays
$$
\sigma[\tau\tau\bar e\bar e\  +~\etslash]:
\sigma[\tau\tau\bar\mu\bar\mu\  +~\etslash]
:\sigma[\tau\tau\bar\mu\bar e\  +~\etslash]:\sigma[\tau\tau (4j)]
= 1:1:2:9.
\eqno (14)
$$
Evidently, the differences between the leptonic signals provide the
cleanest distinction between the neutralino and heavy Majorana neutrino
scenarios.

In case our $LSP$ is different from the neutralino, the $\tau$-number
violating signals are different and other forms of new physics could mimic
those.  For instance, a new charged lepton, mixing dominantly with the
$\tau$, could lead to signals similar to those discussed in Section 3. We
do not discuss these options in detail.

Turning to models where $R_P$ is spontaneously broken [21] via an $SU(2)
\times U(1)$ singlet sneutrino $VEV$, we note that these allow $W^\pm
\tilde Z_1 \tau^\mp$, $Z\tilde W_1^\pm \tau^\mp$ and $Z\tilde Z_1
\nu_\tau$ vertices with strengths given by the corresponding gauge
couplings times some appropriate mixing factors.  Their visible decay
patterns can be both $Z$- and $W$-mediated.  They are $\tilde Z_1
\rightarrow \tau f_u \bar f_d,\bar \tau\bar f_u f_d, \nu_\tau\bar
ff,\bar\nu_\tau f\bar f$ and $\tilde W^-_1 \rightarrow \nu_\tau\bar f_u
f_d, \tau\bar ff$, $\tilde W^+_1 \rightarrow \bar \nu_\tau f_u \bar
f_d,\bar\tau f\bar f$ where $f$ is any fermion while $f_u(f_d)$ is a
fermion of the up (down) type.  Various multilepton and/or multijet final
states with or without $\etslash$ are possible from a $\tilde W^+_1
\tilde W^-_1$ or $\tilde Z_1 \tilde Z_1$ pair.  We see, though, that the
situation is rather similar to that with a decaying heavy lepton (either
neutral Majorana or charged and dominantly mixing with the $\tau$-family)
pair and it would be hard to distinguish between those two scenarios.
However, the tests proposed to distinguish between our explicitly broken
$R_P$ scheme and one with a heavy neutrino can also be used to
discriminate the former from a spontaneously broken $R_P$ model.
Moreover, the presence of a Majoron and an associated light scalar might
lead to additional signatures if $R$-parity is spontaneously violated.
\bigskip

\centerline{\fourteenbf 6. Summary and concluding remarks}

\nobreak
In this paper we have investigated the prospects for detecting explicit
$R_P$-violation at LEP 200 in a $\tau$-number non-conserving scenario. If
$R_P$ is not conserved, a general analysis of supersymmetric signals
becomes very difficult. This is because of the large number of new
interactions that are then possible (see (2)), even assuming that baryon
number is conserved.  However, as reviewed in Section 2, there already
exist experimental constraints on the coupling constants for these new
interactions.  For a sfermion mass $\sim 200$ GeV, we find that $e$- or
$\mu$-number violation can only be substantial (i.e. of electromagnetic
strength) for interactions involving second- or third-generation quarks.
In contrast, rather large $\tau$-number violating couplings are possible
even for purely leptonic interactions, as well as for $\tau$ interactions
with first-generation quarks and squarks.  This is why we have focused on
$\tau$-number violation in our analysis.

Unlike in the MSSM, an unstable $LSP$ need no longer be neutral.  We
have pointed out in Section 1 how any one of the neutralino, sneutrino,
charged slepton,  top squark or  chargino may well be the $LSP$ in an
$R_P$-violating scenario.  As discussed in Sec. 3, for each one of these
cases, the production of $LSP$ pairs at LEP 200 leads to distinctive
signatures in the form of spherical events with $n$ leptons and $m$ jets,
possibly accompanied by a substantial amount of missing energy $(n,m \leq
4)$.  Since we have assumed that $R_P$-violation responsible for $LSP$
decay is simultaneously accompanied by the non-conservation of $\tau$-

number, the final state from the decay of an $LSP$ pair necessarily
contains two leptons from the $\tau$-family.  Our scenario is thus
characterized by the fact that $LSP$ pair production results in
$\tau$-rich final states.  Clearly, the prospects for the detection of
such states will be sensitively dependent on the experimental efficiency
for identifying $\tau$'s.

In order to keep our considerations free from any assumptions about the
masses of other sparticles, we have confined our analysis to signals from
just the production of $LSP$ pairs.  We stress, though, that the production
of heavier sparticles will also lead to $\tau$-rich final states. This is
because those particles can either
decay to the $LSP$ by $R_P$-conserving interactions, or
directly decay to ordinary particles via the $\tau$-number violating
interactions present in our scenario.

As mentioned above, the pair-production of heavy $LSP$s leads to
very distinctive events.  These have been catalogued in Section 3
for each of our $LSP$ candidates.  It is worth emphasizing that, despite
our lack of knowledge about the coupling constants for the $R_P$-violating
interactions, it is possible to relate the cross sections for various
expected characteristic final states.  For the case when the $LSP$ is the
neutralino, these relations are given by (8) - (10) whereas (11) - (13)
are the corresponding relations in the chargino case.

Of the various signals discussed in Sec. 3, most interesting are the
like-sign ditau signals that can result from the production of neutralino
pairs.  First, these are quite spectacular -- especially considering that
the SM backgrounds are tiny.  More importantly, the decays $\tilde Z_1
\rightarrow \tau jj$ and $\tilde Z_1 \rightarrow \bar\tau jj$ lead to
($\tau\tau$ or $\bar\tau\bar\tau$) $+$ $n\le 4$ jet events in which, apart
from measurement errors, any missing $E_T$ arises only from the decays of
the $\tau$, and so tends to be rather soft.  The observation of such
events can potentially lead to unambiguous evidence for $\tau$-number
violation since the smallness of missing $E_T$ makes it unlikely that two
particles carrying $\tau$-number would escape detection in the apparatus.
Of course, detailed studies are necessary before definitive conclusions
can be drawn.  We hope, however, that our somewhat qualitative analysis is
a useful first step.

As can be seen from Figs. 4 and 5, $\sigma(\tilde Z_1\tilde Z_1) \lsim 1$
pb.  We then see from (9) that assuming an integrated luminosity of
500 pb$^{-1}$/yr, about $60~\tau\tau(4j)$ and $\bar\tau\bar\tau$ can be
expected in a year of operation.  Assuming a detection efficiency of 30\%
for terms,  a handful of these spectacular events are possible.
Like-sign ditau events are also possible if a slepton is the $LSP$, though
in this case the total pair production cross section is only about 0.2 pb.

We have, in Section 5, studied rival new physics mechanisms that could
mimic the signals of our $R_P$-violating scenario.  We have shown that
explicit $R$-parity violation can, in principle, be distinguished from
these other new physics mechanisms by studying the ratios of cross
sections for producing various final states.  However, heavy lepton
signals could be confused with those from a spontaneously broken
$R$-parity scenario.

Before closing, we remark that although we have focused our attention on
$\tau$-number violation, it is possible that the dominant $R_P$-violating
operator does not conserve $e$- or $\mu$-number.  This may be because all
of the $\lambda_{ijk}$ and $\lambda'_{ijk}$ are much smaller than their
current bounds [6] discussed in Sec. 2.  Such a situation will, of course,
not affect the signal cross sections since the production mechanism does
not involve these couplings.  Our analysis can easily be carried over to
this case.  In fact, the number of events that could be observed
should then be larger
by a factor of $5 - 10$ from the case of $\tau$-number violation since the
detection efficiency for an $e$ or $\mu$ is considerably larger than that
for a $\tau$.

In summary, we have shown that if $R_P$ is broken by explicit
$\tau$-number violating operators, there are many distinctive signals that
might be observable at LEP 200.  The detectability of these signals
depends crucially on the efficiency of tau identification.  In view of the
novel and promising nature of the new physics, we urge our experimental
colleagues to follow up on these issues.
\bigskip

\centerline{\bf Acknowledgements}

\nobreak
We are grateful to M. Dittmar, C. Gonzalez-Garcia, S. Komamiya, E. Ma,
N.K. Mondal, C. Petridou and D.P. Roy for valuable discussions.
This project was started
during the 2nd Workshop on High Energy Physics Phenomenology (WHEPP II,
Calcutta, January 1991) organized by the S.N. Bose National Centre for
Basic Sciences, and XT thanks the University Research Council of the
University of Hawai and the U.S. National Science Foundation for
financial support during the workshop. RG acknowledges the hospitality of
CERN, PR that of the University of Hawaii and XT that of the Tata
Institute of Fundamental Research and of the Theory Group at KEK. The
research of XT was supported in part by the U.S. Dept. of Energy under
contract no. DE-AM03-76SF00235.
\bigskip

\centerline{\bf Appendix: Model of an unstable heavy doublet Majorana
neutrino}

\nobreak
The simplest model of a heavy Majorana neutrino, $\nu_M$, which contains a
substantial $SU(2)_L$ doublet component and allows for $\tau$-number
violating decays of $\nu_M$, is obtained by adding a sequential
left-handed lepton doublet $\left(\matrix{\ell_4 \cr \nu_4}\right)_L$ and
right-handed singlets $\ell_{4R}$ and $N_R$ to the SM.
Both a Dirac mass $(m_4)$
between $\nu_4$ and $N_R$ Majorana mass $(M)$ for $N_R$ are possible.  In
order to have $\tau$-violating decays, we will also assume a Dirac mass
$(m_3)$ between $\nu_\tau$ and $N_R$.  We will assume that Dirac mass
terms between $\nu_e$ and $\nu_\mu$ and $N_R$ are negligible.  The
 $3 \times 3$ neutrino mass-matrix is

$$
\left(\matrix{0 & 0 & m_3 \cr 0 & 0 &
m_4 \cr m_3 & m_4 & M}\right).
$$

It is easy to see that, apart from the unmixed massless neutrinos $\nu_e$
and $\nu_\mu$, there is another massless state,
$$
(\nu_\tau)_{\rm phys} = m_4 \nu_\tau - m_3 \nu_4.
$$
If $M \gg m_3,m_4$, the two remaining eigenstates have masses $m_{\nu_M} =
m^2_4/M$ and $M$, and are respectively given by
$$
\nu_M = -m_3 \nu_\tau - m_4 \nu_4 + {m^2_4 \over M} N
$$
and
$$
\nu_S = m_3 \nu_\tau + m_4 \nu_4 + M N.
$$
It is straightforward to check that the cross-generation interactions
$W\tau \nu_M$ and $Z(\nu_\tau)_{\rm phys} \nu_M$ are suppressed by factors
of $x = \displaystyle{m_3 \over m_4} $ and
$m^2_4(m^2_4 + m^2_{\nu_M})^{-1}$
$(m_{\nu_M}/M) x$,
respectively.  For natural values, $m_4 = 200$ GeV, $M = 800$ GeV we find
$m_{\nu_M}$ to be 50 GeV.  We will further assume the ratio $m_3/m_4$
to be $\sim 10^{-2}$.  (The smallness of $m_3$ may be speculated to
be due to the smallness of the corresponding intergenerational Yukawa
coupling; this also provides a rationale for neglecting Dirac mass terms
between $\nu_e/\nu_\mu$ and $N_R$).  Then we find that the
intergenerational $W$ and $Z$ interactions are suppressed by $10^{-2}$ and
$\lsim 10^{-3}$.  As a result, there is no conflict between this model and
LEP constraints or data on lepton universality in $W$-decays.  Finally, we
note that the $Z\nu_M\nu_M$ coupling is suppressed by just $m^2_4 (m^2_4 +
m^2_{\nu_M})^{-1}$ which is close to unity so that $\nu_M$ pair production
at LEP is essentially unsuppressed.

\endpage

\centerline{\underbar{\bf References}}

\item{[1]} X. Tata, in {\it The Standard Model and Beyond} (ed. J.E.
Kim, World Scientific, Singapore, 1991), p304. H.P. Nilles, Phys. Rep. {\bf
C110} (1984) 1.  P. Nath, R. Arnowitt and A.H. Chamseddine, {\it Applied
$N=1$ Supergravity} (World Scientific, Singapore, 1984).
H.E. Haber and G.L. Kane, Phys. Rep. {\bf C117} (1985) 75.

\item{[2]} E. Ma and P. Roy, Phys. Rev. {\bf D41} (1990) 988.

\item{[3]} S. Wolfram, Phys. Lett. {\bf 82B} (1979) 65.  C.B. Dover, T.
Gaisser and G. Steigman, Phys. Rev. Lett. {\bf 42} (1979) 1117.  P.F.
Smith et al., Nucl. Phys. {\bf B149} (1979) 525 and {\bf B206} (1982)
333.  E. Norman et al., Phys. Rev. Lett. {\bf 58} (1987) 1403.

\item{[4]} M. Davier, Proc. Intl. Conf. {\it Lepton-Photon
Interactions and Europhys. H.E.P.} (Geneva, 1991, World Scientific,

Singapore, in press), LAL Preprint 91-48.  L. Pondrom, Proc. 25th
Intl. Conf. {\it High Energy Physics} (Vol. 1), p144 (World Scientific,
Singapore, 1991).

\item{[5]} C.S. Aulakh and R.N. Mohapatra, Phys. Lett. {\bf 119B} (1982)
316. L.J. Hall and M. Suzuki, Nucl. Phys. {\bf B231} (1984) 419.  S.
Dawson, Nucl. Phys. {\bf B261} (1985) 297.  S. Dimopoulos and L.J. Hall,
Phys. Lett. {\bf B207} (1987) 210.  L.J. Hall, Mod. Phys. Lett. {\bf A5}
(1990) 467.  R. Arnowitt and P. Nath in {\it Phenomenology of the Standard
Model and Beyond} (eds. D.P. Roy and P. Roy World Scientific, Singapore
1989), p145.  S. Dimopoulos, R. Esmailzadech, L.J. Hall, J-P. Merlo and
J.D. Starkman, Phys. Rev. {\bf D41} (1990) 2099.

\item{[6]} V. Barger, G.F. Giudice and T. Han, Phys. Rev. {\bf D40} (1989)
2987.

\item{[7]} A. Bouquet and P. Salati, Nucl. Phys. {\bf B284} (1987) 557.
B.A. Campbell, S. Davidson, J. Ellis and K. Olive, Phys. Lett.
{\bf B256} (1991) 457.  A. Nelson and S.M. Barr, Phys. Lett. {\bf B258}
(1991) 45. W. Fishler, G.F. Giudice, R.G. Leigh and S. Paban,
{\it ibid}. {\bf B258} (1991) 45.

\item{[8]} H. Dreiner and G.G. Ross, Oxford preprint no. OUTP-92-08P(1992).

\item{[9]} H. Baer et al., in {\it Research Directions for the Decade},
Proc. Snowmass Workshop, 1990 (Editions Frontiere,
Gif-sur-Yvette, in press).  H. Dreiner and G.G. Ross, Nucl. Phys. {\bf
B365} (1991) 597.  H. Dreiner and R.J.N. Phillips, Nucl. Phys. {\bf B367}
(1991) 591.

\item{[10]} A. Nelson and S. Barr, Phys. Lett. {\bf B246} (1990) 141.  H.
Dreiner, Proc. Intl. Conf. {\it Lepton-Photon Interactions and Europhys.
H.E.P.} (Geneva, 1991, World Scientific, Singapore, in press).

\item{[11]} D.P. Roy, Phys. Lett. {\bf B283} (1992) 270.

\item{[12]} D.E. Brahm and L.J. Hall, Phys. Rev. {\bf D40} (1989) 2449.  S.
Lola and J. McCurry, Oxford Preprint OUTP-91-31P, to appear in Nucl. Phys.
B.  H. Dreiner and S. Lola, OUTP 92-02P, to appear in the proceedings of
the EE500 Workshop, Hamburg, September 1991.

\item{[13]} M.C. Bento, L.J. Hall and G.G. Ross, Nucl. Phys. {\bf B292}
(1987) 400.

\item{[14]} L.E. Ib\'a\~nez and G.G. Ross, Nucl. Phys. {\bf B368} (1992) 3.

\item{[15]} M. Chen, C. Dionisi, M. Martinez and X. Tata,
Phys. Rep. {\bf 159} (1988) 201.

\item{[16]} J. Alitti et al., Z. Phys. {\bf C52} (1991) 209; Phys. Lett.
{\bf B280} (1992) 137; ALEPH collaboration, D. Decamp et al., Phys. Lett.
{\bf B265} (1991) 430,475.

\item{[17]} U. Ellwanger, Phys. Lett. {\bf 141B} (1984) 435.

\item{[18]} M. Drees and M. Nojiri, Nucl. Phys. {\bf B369} (1992) 54.

\item{[19]} J. Ellis and S. Rudaz, Phys. Lett. {\bf 128B} (1983) 248.

\item{[20]} J. Ellis and M. Sher, Phys. Lett. {\bf 148B} (1984) 309.  L.J.
Hall and J. Polchinski, Phys. Lett. {\bf 152B} (1984) 335.

\item{[21]} C.S. Aulakh and R.N. Mohapatra, Ref. [5].
A. Santamaria and J.W.F. Valle, Phys. Lett. {\bf B195} (1987) 423;
Phys. Rev. Lett. {\bf 60} (1988) 397; Phys. Rev. {\bf D39} (1989) 1780.
A. Masiero and J.W.F. Valle, Phys. Lett. {\bf B251} (1990) 142.

\item{[22]} J. Carter, Proc. Conf. {\it Lepton-Photon Interactions and
Europhys. H. E. P.} (Geneva, 1991, World Scientific, Singapore, in press).

\item{[23]} S. Dimopoulos and L.J. Hall, Ref. 5.

\item{[24]} H. Baer, M. Drees and X. Tata, Phys. Rev. {\bf D41} (1990)
3414.  J. Ellis, G. Ridolfi and F. Zwirner, Phys. Lett. {\bf B237} (1990)
423.  M. Drees and X. Tata, Phys. Rev. {\bf D43} (1991) 2971.

\item{[25]} H. Baer, V. Barger, D. Karatas and X. Tata, Phys. Rev. {\bf
D36} (1987) 96.

\item{[26]} H. Baer et al. Int. J. Mod. Phys. {\bf A4} (1989) 4111.

\item{[27]} E.A. Paschos and C.T. Hill, Phys. Lett. {\bf B241} (1990) 96.

\endpage

\centerline{\bf Figure captions}


\item{\rm Fig.~1.~:} Purely leptonic $R$-parity and $\tau$-number
violating vertices.

\item{\rm Fig.~2.~:} $R$-parity and $\tau$-number violating vertices
involving quarks.

\item{\rm Fig.~3.~:} Diagrams contributing to four-body decays of $\tilde
\nu_\mu$.

\item{\rm Fig.~4.~:} Cross section contours for neutralino pair-production
$\sigma(\tilde Z_1\tilde Z_1)$, with $\tan\beta \allowbreak = 2$, in the
$2m_1$ -- $m_{\tilde g}$ plane for (a) $m_{\tilde \ell} = 100$ GeV and (b)
$m_{\tilde\ell} = 200$ GeV.  The heavy dotted lines correspond to
$m_{\tilde Z_1} = 45$ GeV.

\item{\rm Fig.~5.~:} Contours for $\sigma(\tilde Z_1\tilde Z_1) = 0.8$ pb,
with $\tan\beta = 1$ (solid) and $\tan\beta = 10$ (dot-dash).  The heavy
circles (triangles) are contours of $m_{\tilde Z_1} = 45$ GeV for
$\tan\beta = 1$ $(\tan\beta = 10)$.

\item{\rm Fig.~6~(a).~:} $p_T$-distributions of electrons and $\tau$'s in
$\tau^+ \tau^+ e^- e^-$ (and c.c.) $+ \etslash$
events from the decays of
neutralino pairs. The normalization assumes that $\tilde Z_1$ dominantly
decays via the $\lambda_{131}$ interaction.
Also
shown are the same distributions from $Z$-pair production, where the
electrons come from the decays of $\tau$'s produced via $Z \rightarrow
\tau^+\tau^-$.

\item{\rm Fig.~6~(b).~:} The distribution of the minimum $\Delta
r(\tau,{\rm jet})$ in $\tau^+\tau^+ (4j)$ events from the production of
neutralino pairs. The normalization is arbitrary.

\end